\documentclass{cambridge6A}
\usepackage{graphicx}
\usepackage{natbib}

\def\be{\begin{equation}}
\def\ee{\end{equation}  }
\def\bea{\begin{eqnarray}}
\def\eea{\end{eqnarray}  }
\def\rg{\sqrt{-g}}
\newcommand{\Screll}{{\mathcal L}}

\begin{document}

\chapterauthor{Luis Lehner and Frans Pretorius}
\chapter{Final State of Gregory-Laflamme Instability}

\contributor{Luis Lehner
\affiliation{Perimeter Institute for Theoretical Physics}
\affiliation{Department of Physics, University of Guelph}
} %

\contributor{Frans Pretorius
\affiliation{Department of Physics, Princeton University}
} %
\begin{abstract}
We describe the behavior of a perturbed 5-dimensional black string
subject to the Gregory-Laflamme instability. 
We show that the horizon evolves in a self-similar manner,
where at any moment in the late-time development of the instability the horizon can be 
described as a sequence of 3-dimensional spherical black holes of varying size, joined by black 
string segments of similar radius. As with the initial black string, each local string segment 
is itself unstable, and this fuels the self-similar cascade
to (classically) arbitrarily small scales; in the process the horizon develops a fractal 
structure. In finite asymptotic time, the remaining string
segments shrink to zero-size, yielding a naked singularity.
Since no fine-tuning is required to excite the instability, this constitutes a
generic violation of cosmic censorship.
We further discuss how this behavior is related to satellite formation in
low-viscosity fluid streams subject to the Rayleigh-Plateau instability, and
estimate the fractal dimension of the horizon prior to formation of the 
naked singularity.
\end{abstract}

\section{Overview}\label{sec_overview}

The ultimate fate of black holes subject to the Gregory-Laflamme instability
has been an open question for almost two decades. In this chapter we discuss
the behavior of an unstable 5D black string and elucidate its final state.
Our studies reveal that the instability unfolds in a self-similar fashion, where the
horizon at any given time can be seen as thin strings connected by hyperspherical black holes of
different radii. As the evolution proceeds, pieces of the string shrink while others
give rise to further spherical black holes, and consequently the horizon
develops a fractal structure. At this stage its overall topology is still $R\times S^2$,
with the fractal geometry arising along $R$ and having an estimated Hausdorff 
dimension $d\approx 1.05$. However, the ever-thinning string-regions
eventually shrink to zero size, revealing a (massless) naked singularity.
Consequently, this spacetime provides a generic counterexample to the cosmic censorship
conjecture, albeit in 5 dimensions.  While we restrict to the 5D case for computational
cost reasons, our observations are intuitively applicable to higher ones.

To correctly capture the late-time nonlinear dynamics of the system
requires numerical solution of the full Einstein equations.
In this chapter, following a brief historical account (Sec.~\ref{sec_history}), we describe
details of our numerical implementation
(Sec.~\ref{approach_chapter3}) as well as the behavior
of the obtained solution (Sec.~\ref{solution_chapter3}). We discuss
some additional properties of the solution, including speculation on when
quantum corrections are expected to become important, and future directions,
in Sec.~\ref{further_chapter3}.

\section{Background}\label{sec_history}

Gregory and Laflamme's observation \citep{Gregory:1993vy}  that linearized perturbations of D-dimensional
black strings ($D \ge 5$) are unstable for long wavelengths is in stark contrast
with the known behavior of black holes in 4D~\citep{1972PhRvD...5.2419P}. Based on the
nature of the growing perturbations of the black strings, together with
entropy arguments, Gregory and Laflamme conjectured black strings would
bifurcate, thus inducing a topology change in the horizon to yield localized $S^{D-2}$
black holes. However, black hole bifurcation necessarily implies, at the classical level,
a formation of a naked singularity where the pinch-off occurs~\citep[see for example][]{Hawking:1973uf}.
Assuming such behavior would be resolved by quantum gravity, the conjecture
was taken as likely true for about a decade. In early 2000s, tension arose when
Horowitz and Maeda proved that any bifurcation could only take place at infinite affine time
along the generators of the horizon that cross the bifurcation point~\citep{Horowitz:2001cz}.
They dismissed this possibility as unlikely and conjectured the existence of stationary,
non-uniform black string solutions which would be the end point of the instability.
Follow up works presented approximate stationary solutions found perturbatively~\citep{Gubser:2001ac} or 
numerically~\citep{Wiseman:2002zc,Kudoh:2003ki},
though these had less entropy than the uniform string and so could not be the end-point of the system.
To these developments, interesting observations were made as to a possible reversal of this behavior
for $D>13$~\citep{Sorkin:2004qq} but that this dimension varies in boosted black strings~\citep{Hovdebo:2006jy}.
Furthermore, it was pointed out
that electrically charged black strings are more unstable than magnetically charged ones~\citep{Sarbach:2004rm},
and arguments were presented for a conical structure in the black string-black hole transition~\citep{Kol:2003ja,Kol:2004ww}.

This flurry of activity not only hinted at the possibility of rich
phenomenology awaiting in the dynamics of the system, but the need for a full analysis to uncover it.
A first attempt to do so was presented in~\citealt{Choptuik:2003qd}. This study revealed
that the development of a black string, perturbed by a long wavelength periodic mode,
progressed to a structure that could
be described as a sequence of $S^3$ black holes
joined by strings, though could not uncover the final fate as
the code was unable to evolve the solution further. Detailed follow-up analysis of the
results from this simulation showed that the affine time along the generators grew faster
than a simple exponential of asymptotic time~\citep{Garfinkle:2004em}, suggesting 
consistency of a possible pinch-off with the
theorem presented in~\citealt{Horowitz:2001cz}~\citep[see also][]{Marolf:2005vn}.

Following these works, additional hints to the possible end state came by analogy with
fluid systems. The membrane paradigm~\citep{Thorne:1986iy} first suggested that event horizon
dynamics could be described, to leading order, by the Navier-Stokes equations for
a viscous fluid (albeit with some ``unusual'' properties, such as a negative
bulk viscosity). More recently, descriptions of black brane dynamics
in asymptotically 5D Anti de-Sitter spacetime~\citep{Bhattacharyya:2008jc},
and an effective worldvolume theory of black holes
~\citep[the ``black folds'' paradigm,][]{Emparan:2009cs,Emparan:2009at}, similarly describe
the dynamics by Navier-Stokes equations. Cardoso and Dias
noted the qualitative similarity between the dispersion relation
of unstable modes of black strings to those of thin fluid streams
subject to the Raleigh-Plateau instability
\footnote{Though note that the relationship is not exact; in particular
in the long wavelength $\lambda$ limit the growth rate $\Omega$ in
Raleigh-Plateau differs from Gregory-Laflamme as $\Omega\propto\lambda^{-1/2}$ vs $\Omega\propto\lambda^{-1}$
\citep[][]{Cardoso:2006ks,Camps:2010br}. However, as the results here show, the behavior of the system
is determined by ``intermediate'' wavelengths, thus the infinitively long wavelength behavior is likely irrelevant as far as the late time behavior of the string is concerned.}, and
pointed out that if the similarities persisted beyond linear development the black string solution could
share the same fate as the fluid, which {\em does} pinch-off~\citep{Cardoso:2006ks}. Moreover, in the latter system
evolution to pinch-off can be accompanied by a phenomena known as satellite formation, where one
or more generations of ever smaller spherical ``beads'' form in the thinning stream;
the lower the viscosity of the fluid, the more generations are observed~\citep[see for example][]{Eggers:1997zz}.
In fact, as we describe here, this {\em is} qualitatively what is seen to occur in the black
string~\citep{Lehner:2010pn}.
The self-similar nature of this solution makes it impossible to numerically follow
the evolution to arbitrary small scales, though following the trends, together with
the fact that there is no intrinsic length scale in the field equations of general relativity,
suggests that infinitely many generations of satellites will form (at the classical level).
Interestingly, the horizon-fluid analogues describe the horizon as a perfect fluid with a shear viscosity to
entropy density ratio $\eta/s$ of $1/4\pi$, implying a lower viscosity than any ``real-world''
fluid~\citep{Kovtun:2004de,Buchel:2003tz}.

\section{Numerical approach} \label{approach_chapter3}
As mentioned, we relied on numerical simulations to reveal the solution of an
unstable black string. In this section we describe particularly relevant aspects
of the numerical implementation adopted. We note however that achieving a reliable
implementation of Einstein's equations for any given class of problem require
dealing with many subtle issues. It is beyond the scope of this chapter to go into the full
details~\citep[for textbook introductions to numerical relativity
see][]{Bona-and-Palenzuela-Luque-2005:numrel-book,Alcubierre:2008m,Baumgarte:2010tw};
however, since there is as of yet
no universal approach to all problems requiring numerical relativity, we find it useful
to describe the method that has been successful in evolving the
5D black string spacetime, focusing on details particular to this problem\footnote{For other
examples of numerical relativity exploring dynamical scenarios in higher dimensional settings
see~\citep{Choptuik:2003qd,Sarbach:2003dz,Guzman:2007ua,Witek:2010qc,Witek:2010az,Shibata:2010wz,Okawa:2011fv}}.

In the class of ``not-so-subtle'' issues for successful solution
of the initial boundary value problem are:
solving the constraint equations for consistent initial data;
using a mathematically well-posed formulation of the field equations
that furthermore does not admit exponentially growing solutions from
truncation-error-seeded constraint violations;
dealing with the geometric singularity inside the black string; and
using a stable numerical integration scheme that can adequately resolve
all the relevant length scales.
The initial data we use here is the same as in the earlier numerical
study~\citep{Choptuik:2003qd}, and we describe the ansatz and solution method in Sec.~\ref{sec:id}.
For evolution, the present study uses the harmonic
decomposition~\citep{Pretorius:2004jg}\footnote{For additional examples of its use in
numerical relativity see, e.g.~\citep{Garfinkle:2001ni,Szilagyi:2002kv,Lindblom:2005qh,Palenzuela:2009yr}}
 with constraint damping~\citep{Gundlach:2005eh,Pretorius:2005gq},
and is described in Sec.~\ref{sec:evo}.
We use the {\em excision} approach to remove the black string singularity
from the computational domain. This relies on having a detailed
description of the apparent horizon of the spacetime, which
is also a key structure employed to analyze the dynamics of the unstable
string. In Sec.~\ref{sec:ah} we describe basic properties of an
apparent horizon and how it is found in the simulation.
For our numerical scheme, we use finite difference techniques with adaptive
mesh refinement (AMR). These methods will not be discussed here,
but where appropriated references to more information will be given.

``Subtle'' issues of import to the 5D black string evolution will be discussed
in the following sections as well, including the coordinates employed and an appropriate
choice of constraint damping parameters.

\subsection{Initial Data}\label{sec:id}
We are interested in studying the black string's dynamics with respect to perturbations in vacuum.
Here, rather than adopting generic perturbations we concentrate on those that only
break the symmetry along the extra ``string'' direction $w$ (henceforth we will refer to it as ``string-like'' direction). As has been shown in the original
work~\citep{Gregory:1993vy,Gregory:1994bj}, perturbations
in the $S^2$ $(\theta,\phi)$ sector lead, at the linear level, to sub-dominant modes with respect to
those in $w$. Thus restricting to such a case should not affect the generality of the observed
behavior and, as an important added bonus, allows us to restrict to a numerical implementation
with only $2$ relevant spatial coordinates.

In what follows we summarize the approach adopted to define such data~\citep[for full
details see][]{Choptuik:2003qd}. To obtain consistent initial data, we adopt
a Cauchy ($4+1$) decomposition of Einstein equations. At an initial ($t=0$) hypersurface,
its intrinsic metric ($\gamma_{ij}$) and extrinsic curvature ($K_{ij}$) provide suitable data
if they satisfy the Hamiltonian and momentum constraints:
\begin{eqnarray}
H &\equiv& {}^{(4)} R + K^2 - K_{ij} K^{ij} = 0 \, ,\\
M_i &\equiv& D_j (K_i^j - \gamma_i^j K) = 0  \, ,
\end{eqnarray}
with ${}^{(4)}R$ the Ricci scalar associated to $\gamma_{ij}$, $K\equiv \gamma^{ij} K_{ij}$ and
$D_i$ the covariant derivative compatible with $\gamma_{ij}$.
We only consider perturbations depending on $(r,w)$, respecting the $SO(3)$ symmetry; then the general intrinsic
metric can be expressed as,
\begin{equation}
{}^{(4)}ds^2=\gamma_{rr} dr^2 + 2 \gamma_{rw} dr dw + \gamma_{ww} dw^2 + r^2 \gamma_{\Omega} d\Omega^2,
\end{equation}
where $d\Omega^2$ is the unit 2-sphere metric.
With these assumptions
$M_{\theta}=M_{\phi}=0$ so the constraints provide three equations for eight variables. Thus three unknowns
can be solved for provided the remainder information is prescribed. To do so, notice that
in ingoing Eddington-Finkelstein coordinates, the unperturbed black string has
\begin{eqnarray}
\gamma_{rr} = 1 + \frac{2M}{r} ~ , ~ \gamma_{rw}=0  ~ &,& ~ \gamma_{ww} = 1 ~ , ~ \gamma_{\Omega}=1\, , \\
K_{rr} = -2M \frac{(r+M)}{r^3} \alpha~ , ~ K_{rw}=0  ~ &,& ~ K_{ww} = 1 ~ , ~ K_{\theta\theta}= 2M  \alpha \, ;
\end{eqnarray}
with $\alpha=\sqrt{r/(r+2M)}$ and $M$ the mass per unit length. We thus adopt $\gamma_{rw}=\gamma_{ww}-1=K_{rw}=K_{ww}=0$ and introduce
the perturbation by defining
\begin{equation}
\gamma_{\Omega} = 1 + A \sin\left(w \frac{2\pi q}{L}\right) e^{-(r-r_o)^2/\delta_r^2} \, .
\end{equation}
Here, $A$ is a parameter controlling the overall strength of the perturbation, $q$ is an integer to
defining its spatial frequency in the $w$ direction, and $r_0$ and $\delta_r$ are parameters
controlling the extent of the perturbation in the radial direction.
For the results presented here $A=0.1,q=1,r_o=2.5$ and
$\delta_r=0.5$. The remaining variables are obtained numerically by solving the Hamiltonian and
momentum constraints. To this end, we adopt a finite difference approximation of the constraints on a 
uniform grid with $(r,w) \in [R_{min},R_{max}]\times[0,L]$ and deal with the constraints as follows.
The Hamiltonian constraint provides an equation for
$\gamma_{rr}$ which can be schematically written as
\begin{equation}
F_1\, \partial_r \gamma_{rr} + F_2\, \gamma_{rr} \partial_{ww} \gamma_{rr} + F_3\, \gamma_{rr} \partial_w \gamma_{rr}
 + F_4\, (\partial_w \gamma_{rr})^2 + F_5\, (\gamma_{rr})^2 + F_6 \gamma_{rr}=0,
\end{equation}
with $F_l$ ($l=1..6$) being functions that do not depend on $\gamma_{rr}$. This equation is integrated outwards
from an inner boundary chosen well inside the unperturbed black string horizon ($R_{min}=M$), and boundary data there is
provided by the unperturbed value of $\gamma_{rr}$. A second order accurate radial integration is adopted
where the presence of the $w-$derivatives make the numerical problem a non-linear, cyclic (due to the $w$-periodicity), tridiagonal system for the unknowns $\gamma_{rr}|_{r_i,w_j}$. This system is solved using
Newton's method and a cyclic tridiagonal linear algorithm.

The momentum constraint along the $r$ direction provides a first order equation for $K_{\theta\theta}$
of the form,
\begin{equation}
G_1 \partial_r K_{\theta\theta} + G_2 K_{\theta\theta} + G_3 = 0 \, ,
\end{equation}
with $G_l$ ($l=1..3$) being functions that do not depend on $K_{\theta\theta}$. This is a simple ODE
equation which is integrated along each $w={\rm const.}$ line, employing second-order finite differences,
and  boundary conditions at $R_{max}$ provided by the unperturbed black string solution.
The momentum constraint along the $w$ direction provides
an equation for $K_{rr}$ of the form
\begin{equation}
H_1 \partial_w K_{rr} + H_2 K_{rr} + H_3 = 0\, ,
\end{equation}
with $H_l$ ($l=1..3$) being functions that do not depend on $K_{rr}$. This is again a simple ODE
that is integrated to second order accuracy along $r={\rm const.}$ lines. Boundary conditions are specified
along $w=w_{min}$ with the unperturbed values.

The procedure above provides consistent data for our problem. We stress that it is by no means a general
procedure to study consistent initial data for a black string problem, rather it is a straightforward
way of setting up a particular perturbation of the black string to study its
subsequent evolution\footnote{For alternatives see~\citep{Wiseman:2002zc,Sorkin:2002nu,Anderson:2005zi}.}.

The method described above provides data for $\{g_{ij},K_{ij}\}$ ($i,j=r,\theta,\phi,w$). However, data
required for the harmonic formulation (discussed in the following sub-section) consists of $\{g_{\mu\nu},g_{\mu\nu,t}\}$.
Note also that for evolution we impose the harmonic condition
with respect to {\em Cartesian} coordinates $(x,y,z,w)$, thus we first transform
$(r,\theta,\phi)$ to $(x,y,z)$ via the standard relations between 3D Cartesian
and spherical polar coordinates.
Data for $g_{\mu\nu}$ can be defined straightforwardly making use of the relation between metrics as provided
by a standard $4+1$ Cauchy decomposition:
\begin{eqnarray}
g_{00} = -\alpha^2 + \gamma_{ij} \beta^i \beta^j ~,~ g_{0i} = \gamma_{ij} \beta^j ~,~ g_{ij} = \gamma_{ij} \, ;
\end{eqnarray}
where we adopt the unperturbed values for the lapse $\alpha$ and shift $\beta^i$.
Data for $g_{\mu\nu,t}$ is obtained by taking a time derivative of the expressions above, making use of the relation between
$K_{ij}$ and $\gamma_{ij,t}$
\begin{equation}
-2 \alpha K_{ij} = (\partial_t - \mathcal{L}_{\beta}) \gamma_{ij},
\end{equation}
and the harmonic coordinate condition, which in $4+1$ form is
\begin{eqnarray}
(\partial_t -\beta^i \partial_i)\alpha &=& -\alpha^2 K \, , \\
(\partial_t -\beta^i \partial_i)\beta^j &=& \alpha \gamma^{jl} ( \alpha \gamma^{mn}\, {}^{(4)}\Gamma_{lmn} - \alpha_{,l}) \, ,
\end{eqnarray}
where ${}^{(4)}\Gamma_{lmn}$ are the Christoffel symbols of the 4-metric.

\subsection{Evolution}\label{sec:evo}

\subsubsection{The Harmonic Decomposition with Constraint Damping}

In the following sub-section we briefly review harmonic evolution, in particular 
with regards to a numerical implementation~\citep[for more details
see][]{Lindblom:2005qh,Pretorius:2007nq}.
{\em Harmonic coordinates} are a set of gauge conditions
that require each spacetime coordinate $x^a$ to independently satisfy the
covariant scalar wave equation:
\be\label{harm_def}
\nabla^\nu \nabla_\nu x^\mu = \frac{1}{\rg}\partial_{\nu}\left(\rg g^{\nu\xi} \partial_{\xi} x^{\mu}\right) \equiv 0,
\ee
where $g$ is the determinant of the spacetime metric $g_{\mu\nu}$.\footnote{{\em Generalized harmonic coordinates}
add an arbitrary set of {\em source functions} $H^a$ to the right hand side
of (\ref{harm_def})~\citep{1985CMaPh.100..525F}; these can be chosen to implement different gauges,
though here we choose $H^{\mu}=0$.}
Harmonic coordinates have a long history in relativity, and are well adapted to describing
black strings, as the metric in harmonic coordinates is regular on the horizon.
More importantly, for numerical evolution, substitution of (\ref{harm_def}) (and its
first covariant gradient) into the vacuum Einstein field equations $R_{\mu\nu}=0$ yields
a system of explicitly symmetric hyperbolic evolution equations for the metric:
\be
\frac{1}{2} g^{\xi\chi}g_{\mu\nu,\xi\chi} + g^{\xi\chi}{}_{(,\mu} g_{\nu)\chi,\xi} + \Gamma^{\xi}_{\nu\chi}\Gamma^{\chi}_{\mu\xi} =0 \label{efe_harm},
\ee
where $\Gamma^{\xi}_{\nu\chi}$ are the Christoffel symbols.
By themselves, these equations admit a larger class of solutions than desired, and it is only the
subset of solutions that satisfy what can now be considered the constraints
\be
C_{\mu}\equiv g_{\mu\nu} \nabla^{\xi} \nabla_{\xi} x^{\nu} = 0 \, ,
\ee
that are of physical interest. The time-derivative of these constraints
can be related to the traditional Hamiltonian and momentum constraints
discussed in Sec.~\ref{sec:id}. At the analytical level, Bianchi identities
imply that initial data (the metric and its first time derivative) satisfying $C_{\mu}=0$
as well as the traditional constraints, will evolve via (\ref{efe_harm}) to a solution where $C_{\mu}=0$
for all time, {\em provided} the boundary conditions are
consistent with $C_{\mu}=0$.

Numerically the situation is more complicated, as truncation error will generically source
non-zero $C_{\mu}$ during evolution, and this typically grows exponentially (even in a convergent
implementation, rendering it difficult to achieve long-time well behaved evolutions). The
cure~\cite[following][]{Gundlach:2005eh}, is to add {\em constraint damping} terms
to the harmonic form of the Einstein equations (\ref{efe_harm}):
\bea
\frac{1}{2} g^{\xi\chi}g_{\mu\nu,\xi\chi} &+& g^{\xi\chi}{}_{(,\mu} g_{\nu)\chi,\xi}  + 
\Gamma^{\xi}_{\nu\chi}\Gamma^{\chi}_{\mu\xi} \nonumber\\
& + & \kappa\left(2 n_{(\mu}C_{\nu)} -(1+\rho) g_{\mu\nu} n^{\chi} C_{\chi}\right)=0 \label{efe_harm_damp},
\eea
where $n^{\mu}$ is a unit time-like vector, here chosen to be the vector normal
to harmonic time $t=$ constant surfaces, and $(\kappa,\rho)$ are the constraint damping parameters.
Notice that the extra terms are homogeneous in $C_{\mu}$, hence when
$C_{\mu}=0$, Eqn. (\ref{efe_harm_damp}) trivially reduces to the Einstein equations.
If $C_{\mu}\neq 0$, perturbation analysis about Minkowski spacetime~\citep{Gundlach:2005eh}
reveals that  for $(\kappa>0,\rho>-1)$, all Fourier modes of $C_{\mu}$ except a zero
wavelength one are exponentially damped. Analytical results are not known
for the efficacy of constraint damping in the strong-field, non-linear
regime, though empirically it has been shown to work for generic
compact binary systems in 4D (involving black holes and neutron stars),
and the unstable black string in 5D. Typically, a value of $\kappa$ that
works well is $\kappa\approx 1/\ell$, where $\ell$ is some characteristic scale
in the problem; here $\kappa$ is set to $1/M$, where $M$ is the initial
mass per unit length of the unperturbed string. In all 4D simulations to date,
$\rho$ has been set to zero; for the black string, a value of $\rho\in(-1..0)$ was {\em essential}
to damp a zero-wavelength mode growing along the string-like dimension $w$. The exact
value however was not too important, and we chose $\rho=-0.5$.

Regarding boundary conditions in the numerical simulation, the domain is
periodically identified in the $w$ direction, and at the outer boundary
Dirichlet conditions are imposed for simplicity, with the metric fixed to the values
of the initial data there. This latter condition is only consistent with $C_{\mu}=0$ to leading order in $1/r_b$,
where $r_b$ is the radius of the outer boundary; hence to avoid any potential
``problems'' that might arise from this, $r_b$ was chosen to be sufficiently large
that the outer boundary is out of causal contact with the string
horizon during the entire length of the simulation.

\subsubsection{Symmetries and the Cartoon Method}

Since the typical computational resources required to numerically
solve a hyperbolic problem scale like $N^D$, where $N$ represents the
number of mesh points needed to resolve a feature of interest
along one dimension, and $D$ is the total number of space-time
dimensions, it would be impossible to solve for a general perturbation
of a 5D (or higher) black string on contemporary computer clusters.
However, as mentioned before, the analysis in~\citep{Gregory:1993vy,Gregory:1994bj}
shows that only modes along the extra string-like direction $w$
are unstable, whereas perturbations within the $S^2$ cross sections
decay exponentially. Hence it is reasonable to expect that one can
obtain a correct picture of the final end-state by restricting
to spherical symmetry within each $w$-constant slice. In other words,
we consider metrics of the form
\be\label{ss_met}
ds^2 = {}^{(3)}g_{\mu\nu} dx^\mu dx^\nu + R^2 \left(d\theta^2 + \sin^2\theta d\phi^2\right),
\ee
where $x^\mu=(x^0,x^1,x^2)=(t,r,w)$ (harmonic time, radial coordinate, string-like direction),
and the three metric ${}^{(3)}g_{\mu\nu}$ and
areal radius $R$ only depend on the 3 coordinates $x^\mu$.

It would seem that the natural way to proceed (as similarly done with the
initial data) is to directly
discretize the field equations with a metric ansatz of the
form (\ref{ss_met}), reducing the problem to a $2+1$ dimensional
simulation. However we were not able obtain long-term stable
evolution doing so within the harmonic formulation
(i.e. requiring that $(t,r,w)$ be harmonic coordinates with appropriate gauge sources).
Possible reasons for the difficulties experienced are related to spherical coordinates
not being harmonic, and the limited set of gauge source functions we tried to account for
this fact did not yield well behaved evolutions. Potentially
related difficulties were reported in studies of harmonic evolution
in 4D spherical symmetry~\citep{Sorkin:2009bc}.

A way around this problem is to consider the full 5D metric
$g_{\mu\nu}$ in harmonic {\em Cartesian} coordinates $(t,x,y,z,w)$ as this
strategy has worked well in 4D. The $S^2$ symmetry of the spacetime
can then be imposed via the corresponding Killing vectors:
\bea
\xi^{\mu}_1 &=& x \left(\frac{\partial}{\partial y}\right)^{\mu} - y \left(\frac{\partial}{\partial x}\right)^{\mu} \, ,\\
\xi^{\mu}_2 &=& y \left(\frac{\partial}{\partial z}\right)^{\mu} - z \left(\frac{\partial}{\partial y}\right)^{\mu} \, ,\\
\xi^{\mu}_3 &=& z \left(\frac{\partial}{\partial x}\right)^{\mu} - x \left(\frac{\partial}{\partial z}\right)^{\mu} \, .
\eea
Thus, a single $(y=0,z=0)$ slice (for example) of the spacetime
is sufficient to reconstruct the entire spacetime by the action
of the Killing vectors on the metric:
\be\label{kill}
\Screll_{\xi_1} g_{\mu\nu} = \Screll_{\xi_2} g_{\mu\nu} = \Screll_{\xi_3} g_{\mu\nu} = 0.
\ee
The practical way to implemented this in the code
is to discretize the given 3D slice of the 5D metric, and then use
the Killing conditions (\ref{kill}) to replace derivatives
orthogonal to the slice (i.e. in the $y$ and $z$ directions)
as required in the harmonic evolution equations (\ref{efe_harm_damp})
with derivatives tangent to the slice (i.e. in the $x$ direction).
For example, expanding $\Screll_{\xi_3} g_{\mu\nu} = 0$ one can
solve for the $z$-gradients of the metric elements as
\be
g_{\mu\nu,z} = \frac{1}{x}\left[z g_{\mu\nu,x} - 2\delta^z{}_{(\mu} g_{\nu)x} + 2 \delta^x{}_{(\mu} g_{\nu)z} \right].
\ee
This is a variant~\citep{Pretorius:2004jg} of the so-called {\em cartoon}
method~\citep{Alcubierre:1999ab}, originally applied to axisymmetric
evolution in 4D spacetime.

\subsection{Apparent Horizons}\label{sec:ah}

Crucial to understanding the dynamics of an unstable black string
is to understand the behavior of its horizon. In a numerical simulation
of finite length, at best one could recover an approximation to the
actual event horizon (EH), for example by looking at the boundary of the
causal past of some region of the spacetime at the last time step
of the simulation. Furthermore, in spacetimes with naked singularities,
as implied by the numerical solution for this system, an event
horizon does not exist. A better local property of the spacetime to study
is the apparent horizon (AH), defined as the outermost marginally outer-trapped
surface (in other words the outermost surface from which the outward null
expansion is exactly zero, and the inward null expansion is
less than or equal to zero at each point on the surface).
Even though apparent horizons are slicing dependent, the earlier
study~\citep{Choptuik:2003qd} showed that the AH was essentially indistinguishable
from the approximate EH everywhere, modulo at late times because of the ambiguity
in defining the region of spacetime whose causal past defines the exterior
of the EH. We thus focus on the AH, which incidentally is also essential
for the excision technique used to remove the geometric singularity inside
the black string from the computational domain.

For more information on how AHs are defined and searched for in a numerical
evolution, see~\citep{Thornburg:2006zb}; we use a {\em flow method} in this code.

\subsection{Evolution Code}
The basics of the numerical evolution code employed can be briefly summarized
as follows.
The harmonic equations with constraint damping (\ref{efe_harm_damp}),
reduced to first order in time by introducing the auxiliary
variables $\dot{g}_{\mu\nu}\equiv\partial_t g_{\mu\nu}$,
are discretized using $4^{th}$ order finite difference methods
with AMR \citep{Berger:1984zza}, as implemented in the
{\tt PAMR/AMRD}~\footnote{\tt http://laplace.physics.ubc.ca/Group/Software.html}
libraries (which also handle parallelization via MPI).
Time integration is via $4^{th}$ order Runge-Kutta. Standard, centered
spatial difference operators are used in the interior
of the grid, while at the inner excision
boundary (chosen to match the shape of the AH, but a
fraction between $10$ and $50\%$ smaller in radius), centered
difference operators are replaced with sideways operators
as appropriate. A $6^{th}$ order Kreiss-Oliger style dissipation
filter is used to control high-frequency truncation error~\citep{Calabrese:2003vx}.

\section{Evolution of an Unstable, 5D Black String} \label{solution_chapter3}
With the previously described implementation, we now concentrate on describing
the results from the evolution of an unstable black string.
We concentrate on a single case where the unperturbed string has periodicity length $20M$.
Such an identification satisfies $L>L_c$, where $L_c$ is the critical wavelength
beyond which modes become unstable, yet only allows a single unstable mode initially.
To study this case, we adopt a computational domain
$(r,w)\in ([0,320M]\times[0,20M])$ (in the following for convenience
we relabel the $x$ Cartesian coordinate as $r$, though of course
$x=r$ on the $y=z=0$ slice of the spacetime). We adopt an initial grid that uniformly covers
the entire domain with $(N_r,N_w)=(1025,9)$ points. As the evolution proceeds
and finer structure arises, the AMR algorithm introduces additional
higher resolution grids where needed, based on a specified
maximum truncation error tolerance.

A typical ``low'' resolution run has a coarsest radial mesh
spacing of $\sim 0.3/M$, which is the resolution at the outer boundary.
Additional levels are added by the AMR algorithm (each refined with a $2:1$ ratio,
in both $r$ and $w$, of the
parent level)
to resolve the region of spacetime near the AH of the string.
Initially $3$ additional levels are added, growing to $16$ additional
levels by the time the simulation was terminated at $t\sim229.0M$ (due
to the prohibitive computational costs that would have been required
to continue).
``Medium'' and ``high'' resolutions, for convergence and error estimates, were specified
by decreasing the maximum allowed (estimated) truncation error by a factor
of 8 and 64 respectively. Due to lengthier run times with increased resolution,
the medium and higher resolution cases were not run for as long. Specifically,
the longest medium (high) resolution case was terminated at
$t\sim 226.8M$ ($t\sim 217.9M$); the high resolution run required around
$100,000$ CPU hours on the {\tt woodhen} cluster at Princeton
University~\footnote{\tt http://www.princeton.edu/researchcomputing/computational-hardware/}.
This translated into roughly $2$ months of essentially continuous running on $100$ processors.

\subsection{Apparent Horizon Dynamics}

To help understand the dynamics of the system, we monitor several relevant quantities, in
particular, the apparent horizon radius $R_{AH}(t,w)$, the total horizon area $A(t)$,
its intrinsic geometry via a flat-space embedding diagram, as well as
two spacetime curvature invariants $I=R_{\mu\nu\chi\xi} R^{\mu\nu\chi\xi}$ and
$J=R_{\mu\nu\chi\xi} R^{\chi\xi\eta\sigma} R_{\eta\sigma}{}^{\mu\nu}$ evaluated on the horizon.
For the curvature invariants, we find it useful to rescale them as,
\begin{equation}\label{K_inv}
K=I R_{AH}^4/12, \ \ \ S=27 \left( 12 J^2 I^{-3}-1 \right ) + 1  \, ,
\end{equation}
as such rescaling yields $K=S=6$ for an $S^3$ Tangherlini black hole while $K=S=1$ for a uniform
black string.

\begin{figure}
\begin{center}
\includegraphics[width=4.0in,clip=true]{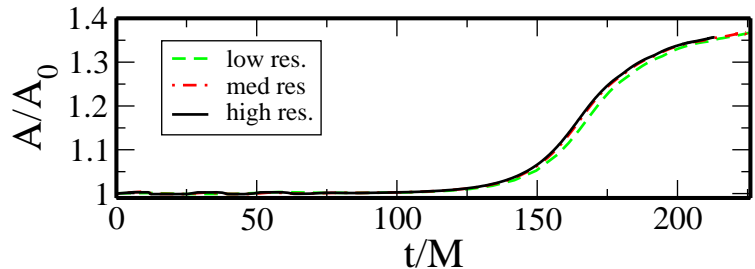}
\caption{Apparent horizon area as a function of time for
the perturbed black string. What is shown is the area normalized
to the initial area, and from simulations with 3 different
resolutions (see the text for a description of the resolutions).
}
\label{fig:AH_radius}
\end{center}
\end{figure}

Fig.~\ref{fig:AH_radius} shows the total apparent horizon area $A$ as a function of time
for the evolution of our perturbed black string.
As expected for a reasonable approximation to the event horizon, the area is
non-decreasing with time. More interestingly, we note that
at the end of the simulation corresponding to the lowest resolution run (the one that ran the farthest in
time) the total area is $A=(1.369\pm0.005) A_0, $\footnote{The error in the area was estimated
from convergence at the latest time data was available from all simulations.} where $A_0$ is the initial area;  this value essentially saturates the value of $1.374 A_0$ that an exact 5D black hole
of the same total mass would have.

\begin{figure}
\begin{center}
\includegraphics[width=4.0in,clip=true]{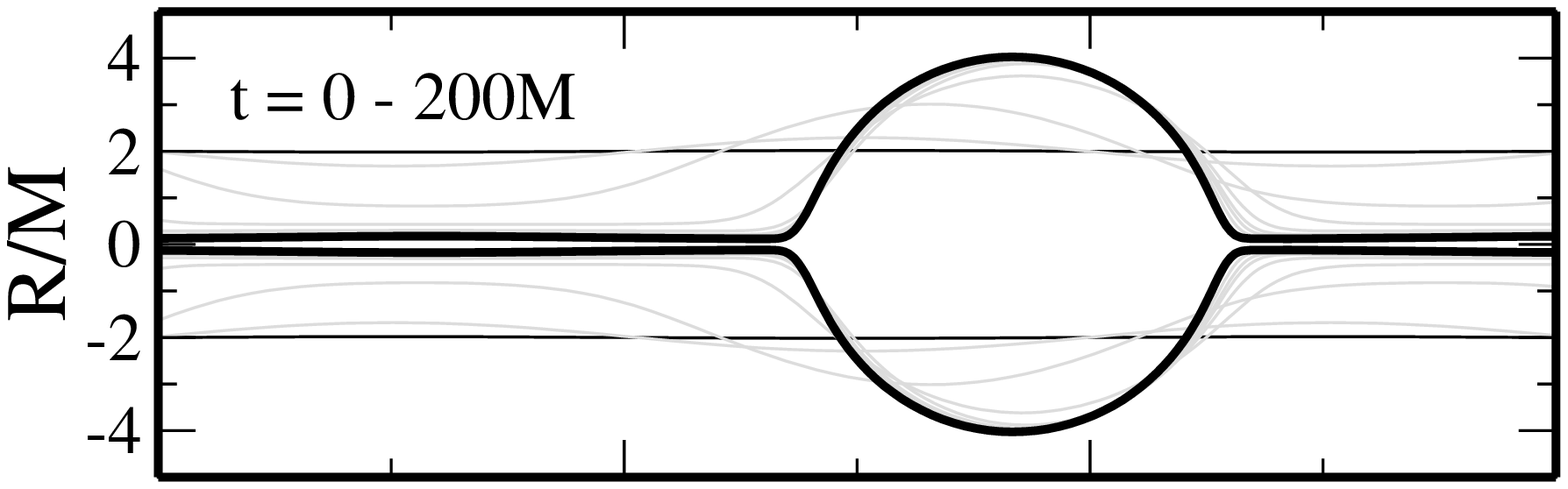}
\includegraphics[width=4.0in,clip=true]{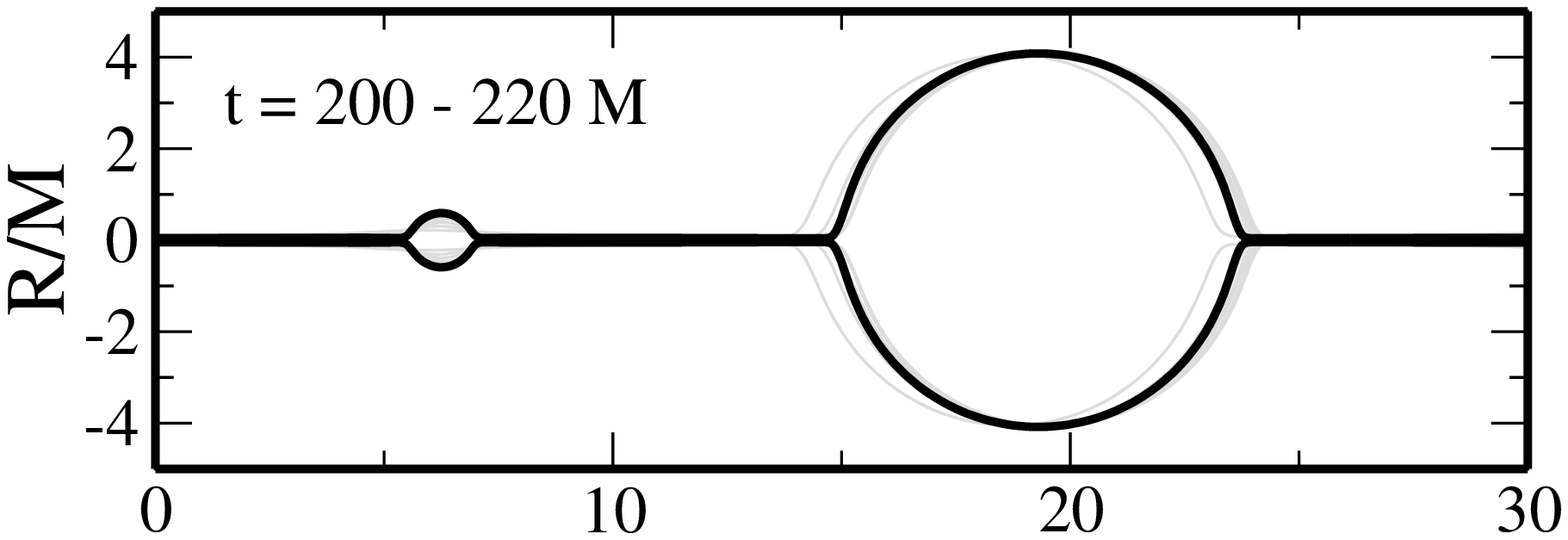}
\includegraphics[width=4.0in,clip=true]{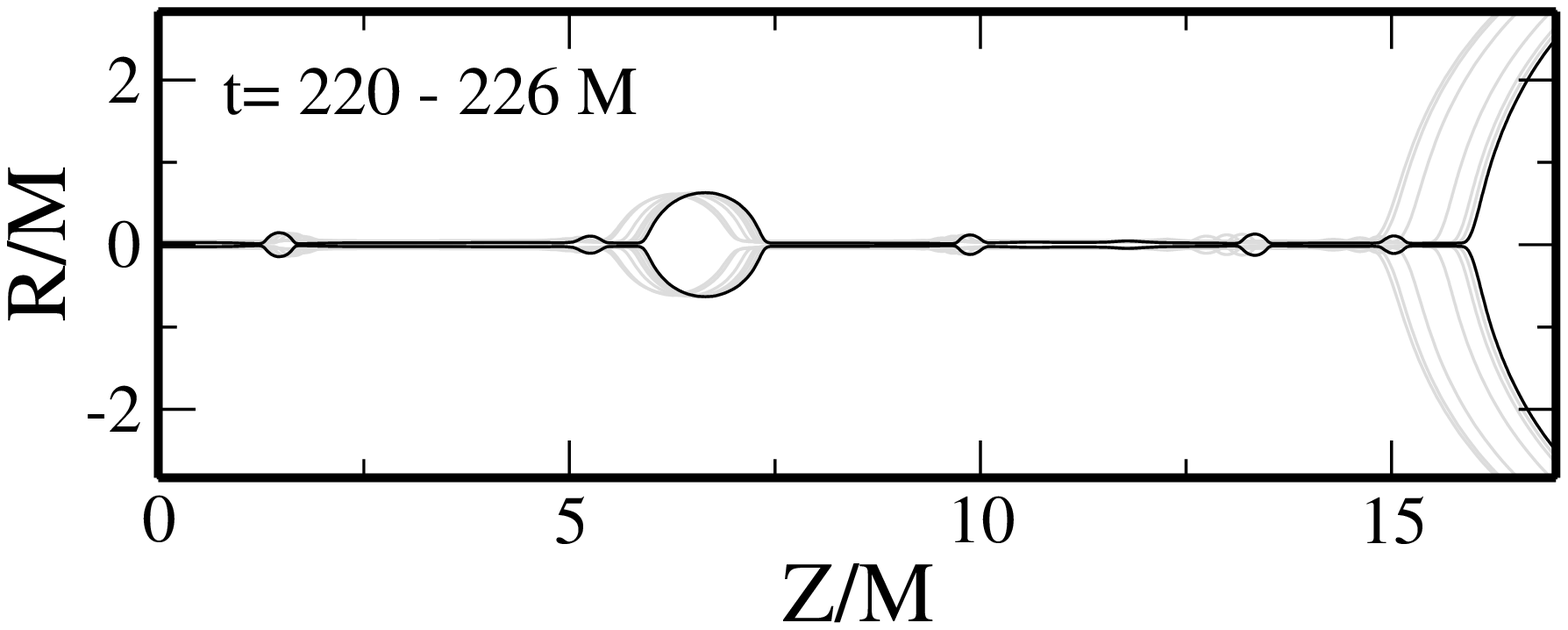}
\caption{Embedding diagram of the apparent horizon at several
instances in the evolution of the perturbed black string,
from the medium resolution run.
$R$ is areal radius, and the embedding coordinate
$Z$ is defined so that the proper length of the horizon in
the space-time $w$ direction (for a fixed $t,\theta,\phi$) is exactly equal to the
Euclidean length of $R(Z)$ in the above figure.
For visual aid copies of the diagrams reflected about $R=0$ have
also been drawn in.
In the top panel, the thin black lines are from the initial time; in the
top two panels the thick black lines are the last time from the time-segment
depicted in the corresponding panel, whereas in the bottom panel
the last time-snapshot has been drawn with a thin black line.
Note that the vertical and horizon axis scales have been changed in the bottom panel
to better show the satellites that have formed at late times.
The computational domain is periodic in $w$ with period $\delta w = 20M$; at  the
initial (final stage of the simulation)  $\delta Z=20M$, ($\delta Z=27.2M$)---see also
Fig.~\ref{fig:AH_length}}
\label{fig:AH_embed}
\end{center}
\end{figure}

More insight into the dynamics of the horizon can be garnered
by observing the evolution of its intrinsic geometry with time. Fig.~\ref{fig:AH_embed} shows
several snapshots of embedding diagrams of the AH, from the medium resolution simulation run.
As can be seen in the figure, the string initially evolves
to a configuration resembling a hyperspherical black hole connected by a thin
string segment, as reported earlier in~\citep[][where the simulation
ended at roughly $t\sim164M$]{Choptuik:2003qd}. However, as the evolution
proceeds it is apparent that the string segments are themselves unstable, and this
pattern repeats in a self-similar manner to ever smaller scales.

\begin{figure}
\begin{center}
\includegraphics[width=4.2in,clip=true]{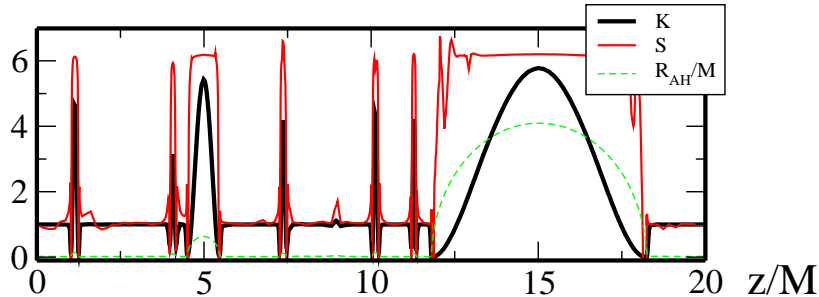}
\caption{Curvature invariants evaluated on the apparent horizon at the
last time of the medium resolution simulation (as depicted in Fig.~\ref{fig:AH_embed}).
Also, for visual aid, the areal radius of the
apparent horizon is shown. The invariant $K$
evaluates to $1$ for an exact black string, and $6$ for an exact
spherical black hole; similarly for  $S$ (\ref{K_inv}).
}
\label{fig:AH_invariants}
\end{center}
\end{figure}

Though the intrinsic geometry on the horizon resembles spheres
connected by strings, this by itself does not imply the local
{\em spacetime} geometry is similar to either 5D spherical black holes
or black strings, respectively. However, evaluation of the curvature
invariants (\ref{K_inv}) on the horizon gives further insight
into this question, and this is shown in Fig.~\ref{fig:AH_invariants}, taken from the last time
step of the medium resolution run. As seen in the figure, near spherical-like sections
the normalized invariants approach $6$, in contrast to the string-like sections
where they are close to $1$; thus, at least based on these two invariant
indicators, the near horizon geometry indeed resembles
that of the solutions suggested by the shape of the AH.

\subsection{Interpretation of the horizon dynamics}

The shape of the AH in the embedding diagram, and that the invariants are
tending to the limits associated with pure black strings or black
holes at corresponding locations on the AH, suggests it is
reasonable to describe the local geometry as being similar
to a sequence of black holes connected by black strings.
This also strongly suggests that satellite formation will continue
in a self-similar cascade, as each string segment locally resembles
a rather uniform black string, and is sufficiently thin and long to be unstable.
Note that even if at some point in the cascade thick segments were to form,
this would generically not be a stable configuration, as (except
possibly in an exactly symmetric situation) the satellites will
have some non-zero $w$-velocity; hence they would eventually merge,
effectively lengthening the segments connecting the remaining satellites.

With this interpretation, we can understand key features of the AH dynamics
by sequencing its time evolution into generations of satellite formation.
We assign a time $t_i$ to the onset of a new generation when the local
instability has reached an ``observable'' level, {\em defined} here (somewhat arbitrarily)
as the time when a nascent spherical region reaches an areal radius $1.5$ times
the surrounding string radius. For each generation we measure the
number $n_s$ of satellite black holes that form per string-segment, their
radii $R_{AH,i}$ as well as the corresponding string segment's radii $R_{s,i}$ and
lengths $L_{s,i}$. These features are summarized in Table~\ref{tab_properties},
where error bars, where appropriate, come from convergence calculations\footnote{The
global truncation error tends to grow with time; also, as mentioned, for the 3rd generation
we only ran the low and medium resolution cases far enough to uncover it, hence (assuming convergence)
the much larger errors there. Only the low resolution run was continued for long
enough to reveal a 4th generation, thus the lack of error-estimates there.}.
Only the first generation is in a sense
non-generic, as, by construction, only one unstable mode is initially allowed to develop.
Generations after the first one
display string segments where their ratio $L_{s,i}/R_{s,i}\simeq 100$ is not only
above the critical one ($\simeq 7.2$), but can in principle accommodate several
unstable modes (though see the discussion in Sec.~\ref{sec_modes}). One would expect the mode closest to the
maximum in the dispersion relation to dominate in each segment, and thus
{\em qualitatively} similar dynamics to unfold from one generation to the next.
In particular, we see that the time scale for the development of the $n^{th}$ generation of
the instability is essentially the same along each string segment, as are the
radii of the corresponding strings and spheres that form. However,
beyond the second generation we find variation in the number $n_s$ of satellites
that form as a function of resolution (hence the $\geq 1$ in the table).
This suggests that exactly which mode
dominates depends sensitively on the initial conditions, sufficiently so
that a small perturbation, here coming from numerical truncation error\footnote{In slight
abuse of terminology, as of course truncation error does not in general correspond
to any physical perturbation.}, can change
the {\em quantitative} details of the late time dynamics. This is not unexpected,
as from the Gregory-Laflamme dispersion relation, except for the mode exactly
at the maximum, there are two unstable modes with the same growth rate.

\begin{table}
{\small
\begin{tabular}[t]{| c || c | c | c | c | c |}
\hline
 Gen. & $t_i/M$ & $n_s$  & $R_{s,i}/M$ & $R_{AH,f}/M$ & $L_{s,i}/R_{s,i}$\\
\hline
 1 & $118.1\pm0.5$ & $ 1 $   & $2.00$        & $4.09\pm0.5\%$ & $10.0$ \\
\hline
 2 & $203.1\pm0.5$ & $ 1 $   & $0.148\pm1\%$ & $0.63\pm2\%$   & $105\pm1\%$ \\
\hline
 3 & $223\pm2$     & $ \geq1$   & $0.05\pm20\%$ & $0.1 - 0.2$    & $\approx  10^2$ \\
\hline
 4 & $\approx 227$   & $ \geq1 $& $\approx 0.02 $ & ?          & $\approx  10^2$  \\
\hline
\end{tabular}
}
\caption{Properties of the black string apparent horizon (see the text
for a discussion).}
\label{tab_properties}
\end{table}

\subsubsection{Extrapolation to the end-state}

The above observation of solution properties allow us to extrapolate
to the end-state of the instability, estimating the time when this will
occur, and the structure of the horizon just prior to it. We first calculate
when the self-similar cascade will end, namely, the time when the connecting string segments reach
zero radius, and then estimate the fractal dimension of the AH geometry
just prior to this end.

The time when the first generation
satellite appears is controlled by the perturbation imparted by the initial data, which here
is $T_0/M\approx 118$. Subsequent generations, however should represent the generic development
of the instability.
From the data in the table, the time of growth of the first
instability {\em beyond} the one sourced by the initial data
is $T_1/M\approx 80$. Beyond that, with the caveats that we have
a small number of data points and poor control over the error at late
times, the data {\em suggests} each subsequent instability unfolds
on a time-scale $X\approx1/4$ times that of the preceding one.
Again this is to be expected if the instability is qualitatively like
the Gregory-Laflamme instability of the exact black string, where the time
scale is proportional to the string radius. Then, the total time $\Delta T$
for the end-state to be reached is roughly
\be
\Delta T \approx T_0 + \sum_{i=0}^\infty T_1 X^i = T_0 + \frac{T_1}{1-X}.
\ee
For this case then, $\Delta T /M\approx 231$. At this time\footnote{The exact value
of which could be expected to vary slightly along the length of the horizon
with our chosen time coordinate and initial perturbation, though one should
be able to define a time-slicing where this time is exactly the
same everywhere along the string.}
all string segments will reach zero radius. Since
the Kretchman curvature invariant (\ref{K_inv}) just outside a string segment of radius
$r$ is proportional to $r^{-4}$, and recalling that here (harmonic) time
is regular everywhere from some distance inside the AH outwards and corresponds
to the time measured by stationary asymptotic observers, this indicates {\em formation of a naked singularity, and
thus a violation of cosmic censorship}. Furthermore, it
is generic in the sense that no fine-tuning of the initial data
is required (i.e., {\em any} length-wise perturbation exceeding the critical wavelength will do).

\begin{figure}
\begin{center}
\includegraphics[width=4.0in,clip=true]{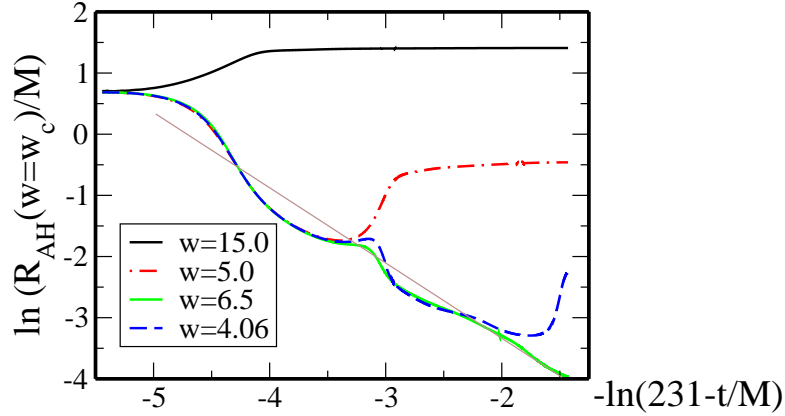}
\caption{
Logarithm of the areal  radius vs. logarithm of time
for select points on the apparent horizon,
from the same (medium) resolution run depicted in Fig.~\ref{fig:AH_embed}.
We have shifted the time axis {\em assuming} self-similar behavior;
the putative naked singularity forms at asymptotic time $t/M\approx 231$.
The coordinates
at $w=15,5$ and $4.06$ correspond to the maxima of the areal
radii of the first and second generation satellites, and
one of the third generation satellites at the time the simulation
stopped (the $w$ positions of the satellites do slightly evolve with time).
The value $w=6.5$ is a representative slice that is in the middle of a
piece of the horizon that remains string-like throughout the evolution.
For visual aid in comparison with the analogue fluid pinch-off scaling
solution (\ref{ns_scaling}), a line with slope equal to -1 has been
added (thin brown line).
}
\label{fig:AH_radius_vs_lnt}
\end{center}
\end{figure}

To give further evidence that the data supports the above conclusion,
in Fig.~\ref{fig:AH_radius_vs_lnt} we show the time evolution
of the radius of several representative cross-sections of the AH, in logarithmic
coordinates where time has been shifted by the estimated time
of naked singularity formation. That the radius
of string-like segments decrease {\em linearly} in shifted-logarithmic time
is consistent with a self-similar scaling to zero radius at the corresponding
finite asymptotic time. Another intriguing aspect of this self-similar scaling
is that it is qualitatively the same as that observed in the approach to
pinch-off of Raleigh-Plateau unstable fluid streams. In the fluid case,
a scaling solution is known~\citep{1993PhRvL..71.3458E,Miyamoto:2010ga},
and the radius $R$ of the fluid column decreases linearly with time $t$ to
fluid breakup $t=t_c$:
\begin{equation}\label{ns_scaling}
R \propto (t_c-t).
\end{equation}
Fig.~\ref{fig:AH_radius_vs_lnt} shows that this, to good approximation,
also describes shrinking regions of the black string.

\begin{figure}
\begin{center}
\includegraphics[width=4.0in,clip=true]{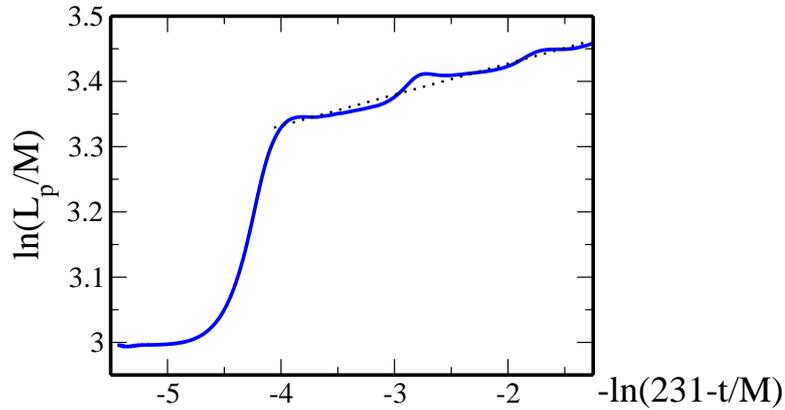}
\caption{The proper length $L_p(t)$ vs. time of the apparent horizon
in logarithmic coordinates, along a fixed azimuthal
cross-section ($\theta=\phi=$ constant in spherical coordinates) and
over one period of the spacetime in the $w$ direction.
The time coordinate is scaled as in Fig.\ref{fig:AH_radius_vs_lnt}.
The dotted line is a visual linear fit to
the late-time dynamics, and has a slope of $\sim0.048$.
\label{fig:AH_length}
}
\end{center}
\end{figure}

A further consequence of the self-similar nature of the instability
is that the AH shape will develop a fractal structure
prior to pinch-off. This implies that for a fixed angular cross section ($\theta,\phi={\rm const.}$)
of the horizon, the proper length $L_p(t)$ (within the periodically identified domain) will
grow with each subsequent generation, diverging at a rate related to the Hausdorff dimension
$d$ of the end-state shape. Assuming a scaling relation of the form~(\ref{ns_scaling}),
that the additional length in each new generation is caused by every string segment developing
(on average) the same number $n_s$ of satellites, each with radii $R_{AH,f}$ following the scaling
suggested in Table~\ref{tab_properties}, and that $n_s \times R_{AH,f}/L_{s,i}\ll 1$, one can show
that the following growth of $L_p(t)$ is expected
\begin{equation}
L_p(t) \propto (t_c-t)^{(1-d)}.
\end{equation}
Figure~\ref{fig:AH_length} shows a plot of $L_p(t)$
on a logarithmic scale. From the measured slope and above relationship,
we find $d \simeq 1.05$, which as expected is greater than a value of $1$
(that would correspond to a non-fractal curve) but only slightly, as the fractal
structure is obtained by repeatedly replacing a relatively long straight line
with a similar length line plus a small semi-circular protrusion.

\section{Speculations and Open Questions} \label{further_chapter3}

In this chapter we described the evolution of an unstable black string in 5D,
at least as far as was numerically feasible given finite computational resources, and extrapolated
the observed behavior to describe the nature of the end-state of the instability.
This has provided a first insight into the fascinating dynamics of unstable
horizons in higher dimensions, yet has raised many questions and opened up avenues
for future work. In this section we conclude by discussing in Sec.~\ref{sec_modes} 
whether the Gregory-Laflamme linear perturbative analysis can be used to quantitatively
understand the dynamics of subsequent generations of the string instability, in Sec.~\ref{sec_quant}
mention when quantum corrections are expected to become important as the string thins, and end
with a list of open questions for future directions in Sec.~\ref{sec_future}.

\subsection{Mode behavior}\label{sec_modes}

\begin{figure}
\begin{center}
\includegraphics[width=3.0in,clip=true]{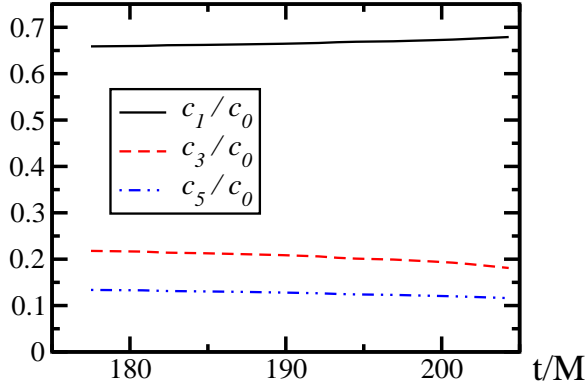}
\caption{Vales of expansion coefficients for the string segment corresponding
to the first generation. Even modes (except $c_0$) are a few orders of magnitude smaller
than odd modes.}
\label{fig:fourier}
\end{center}
\end{figure}

To gain more insight into what determines the observed dynamical behavior,
we analyze the different modes in a string segment at the late stages
of the first generation shown in Table~\ref{tab_properties}, corresponding
to the early stages of the growth of the second generation.
In particular, we look at $R_{AH}(t,w)$ in $t\in(175M,205M)$ when only one
spherical region is clearly distinguishable, and, {\em in our coordinates} this region
has extent in $w\in[10,20M]$. The string section corresponds to $w\in[0,10M]$, and
we decompose $R_{AH}(t,w)$ in this latter domain via the following expansion
\begin{equation}
R_{AH}(t,w) = c_0 - \Sigma_{l=1}^{\infty} c_l \sin(l \pi w)\, ,
\end{equation} 
and extract the coefficients $c_l$ up to $l=6$. Figure \ref{fig:fourier} illustrates the
values $c_l/c_0$ within this time frame, and we only plot the odd values of $l$ as $\{c_{2},c_{4},c_{6}\}$ 
are about two orders of magnitude smaller. This is consistent with the observed symmetry 
in the development of the single second generation hypersphere within this segment, as
it forms exactly in the middle of the string. Interestingly, of the odd modes, only $l=1$ displays
growing behavior even though the length/radius ratio of the string at this time would
allow for several modes to be unstable according to the Gregory-Laflamme instability criteria. This may not
be too surprising however, as during this stage the solution is highly dynamical, shrinking in radius
at a rate comparable to that at which putative Gregory-Laflamme modes could grow. This suggests that 
the linear analysis perturbing about the static black string ``background'' is not applicable here, 
and can only 
capture the qualitative nature of subsequent generations of instability.
Note though that this is not in tension with the observation provided by the invariants $K$ and $S$, as the string
can shrink in a time-dependent fashion and still maintain $K\approx S~\approx 1$. 

\subsection{Dynamics Beyond the Classical Regime}\label{sec_quant}

As discussed, our {\em classical} description of the system reveals cascading
 self-similar behavior of the horizon, with regions evolving to ever shrinking cross-sectional
radius. Such regions asymptote (in finite time) to a zero-mass naked singularity.
Obviously for small enough regions, a classical description is no longer
applicable and quantum phenomena must be taken into account. This is typically
expected to happen when the cross-sectional radius $r$ approaches some small
length $\ell$ scale (e.g. the Planck length $l_p$ or the string length $l_s$),
though exactly when depends on the details of the leading order corrections
to the field equations. For example, stringy-corrections predict this to be of the
form
\begin{equation}\label{r_correct}
R_{\mu\nu} = \ell^2 R_{\mu\alpha\beta\gamma} R_{\mu}{ }^{\alpha\beta\gamma} + \cdots  .
\end{equation}
Since each term in (\ref{r_correct})
has dimension 1/length$^2$,
a naive estimate of when the RHS becomes important is when its magnitude becomes
of order $1/r^2$, as $r$ is the relevant length
scale in the vicinity of a shrinking region of the horizon.
Since $R_{\mu\alpha\beta\gamma}\sim 1/r^2$ there, this suggests
corrections to general relativity are important 
when $\ell^2/r^4\sim 1/r^2$; i.e. if $\ell=l_p$, when the
radius shrinks to the Planck length.

Another effect that could in principle alter the classical picture before
the Planck scale is reached is if the timescale of Hawking evaporation $\tau_H$
of the string segment becomes smaller than the Gregory-Laflamme timescale
$\tau_{GL}$. However, as the following shows, $\tau_H$ only becomes
smaller than $\tau_{GL}$ beyond the Planck scale.

Assuming that a 4D description of evaporation at a $w={\rm const.}$ cross section of a
string-like region gives a decent approximation to an evaporating string,
we can estimate the timescale of Hawking evaporation
and compare it to the timescale for development of the next generation within
the self-similar cascade.
From the simulation results, we have roughly that
\begin{equation}
\tau_{GL} \simeq 100 M_i G/c^3,
\end{equation}
while that for Hawking evaporation of a 4D black hole of mass $M_i$ is
\begin{equation}
\tau_H = 10240 \pi^2 \left( \frac{c^5}{h G} \right) \left( \frac{G M_i}{c^3} \right)^3.
\end{equation}
Thus, for these two times to become comparable
\begin{equation}
M_i \simeq 10^{-1} \sqrt{\frac{\hbar c}{G}} = 10^{-1} M_P,
\end{equation}
where $M_P$ is the Planck mass.
This suggests that if the length scale $\ell$ in (\ref{r_correct}) is of
order the Planck length or larger, then higher curvature corrections will generically
be more important in describing the leading order alterations to the classical description
of the instability than Hawking evaporation.

\subsection{Future work}\label{sec_future}

We conclude this chapter by listing some open questions and directions
for future work, though this is by no means an exhaustive list.

First, consider a broader class of initial perturbations of the
5D black string where other unstable modes, and possibly linear combinations of modes
with similar growth rates, are exited. We do not expect anything to change in a radical
fashion, as the qualitative nature of the linear growth of all these modes is similar. 
Nevertheless, a quantitative analysis of such scenarios would tell if this
expectation bares out, and allow one to understand how (if at all) the initial perturbation affects 
the unfolding structure of subsequent generations of the instability. 
On a similar vein, exploring how adding angular momentum to the string changes the picture
would be interesting. Again the qualitative picture should not change, since as argued in~\citealt{Emparan:2009at}, and shown at the linear level 
in~\citealt{Dias:2010eu}, rotation does not suppress the instability.
However in~\citealt{Marolf:2004fya} it was suggested that rotation could induce super-radiant and 
gyrating instabilities; this should
make for much richer dynamics in the approach to the end-state. Allowing for
angular momentum would require less symmetry, making for more expensive numerical evolution.

Second, extract more details about the spacetime, such as the evolution 
of the generators of the horizon~\citep[as in][]{Garfinkle:2004em}, and the gravitational waves that
are emitted as the instability unfolds. Perhaps the most profound question raised by these results is the striking qualitative 
similarity between the {\em non-linear} evolution of the instability and the Rayleigh-Plateau instability.
Is this a coincidence, or the consequence of a deeper relationship between Einstein and Navier-Stokes
than already suggested by the membrane paradigm, black folds and other perturbative descriptions
of horizon dynamics? What would be useful in trying to understand this is 
to identify and extract geometric characteristics of the dynamical horizon that would map to effective 
fluid properties. Such a map could also prove useful in taking what is known about
the Rayleigh-Plateau instability to learn more about Gregory-Laflamme; for example, 
translating the Eggers scaling solution~\citep[][]{1993PhRvL..71.3458E} to an approximate
solution for a thinning string segment.

Third, consider non-vacuum spacetimes;
in particular charged black strings would be interesting as charges modify the onset of
instabilities in the system ~\citep[for e.g.][]{Gregory:1994bj,Sarbach:2004rm}.

Fourth, study a case where the spacetime is asymptotically flat in all spatial
directions (the current study imposed periodicity in the fifth dimension).
This would be required to provide an example of cosmic censorship violation
in the 5D asymptotically flat case. However, it is natural to expect the same behavior as seen
here based on the following putative scenario. Begin with a highly distorted $S^3$ horizon, namely one
which is long and thin (``cigar shaped'') so that near the center of the horizon it
locally resembles an $S^2\times R$ black string. One may expect that this horizon
would ring-down via gravitational wave emission to a uniform $S^3$ horizon; however, if the dynamical
time-scale for the ring-down (which will be proportional to the light-crossing time in the
prolate direction, so could be made arbitrarily long)
is much longer that the Gregory-Laflamme instability of the
central region, then the latter should take over first, resulting in a pinch-off. 

Fifth, explore black strings in higher dimensional spacetimes, particularly to investigate
the conjecture of~\citep{Sorkin:2004qq} that a qualitatively different end-state
is expected for $D>13$. If one continues to impose 
$SO(D-2)$ symmetry the problem can still be expressed to depend only on $(t,r,w)$, making it
tractable with current computational resources.  However, as the fields 
decay as $\{r^{D-3}, r^{D-2}\}$ moving away from black hole, black string solutions respectively, the resolution
and/or order of the numerical scheme would need to be higher than the one used in this work
to obtain results of comparable accuracy. 

Sixth, explore additional black objects subject to Gregory-Laflamme-like instabilities. For instance, rapidly
rotating black holes have been shown to be subject to a similar instability~\citep{Emparan:2003sy}. Recent numerical
work presented the first exploration of such systems~\citep{Shibata:2010wz}, though there
radiation of angular momentum stabilizes the black holes considered. However, the growth rate
of the instability in the cases studied was rather mild, and it is quite likely that choosing
a more extreme scenario (where the timescale of the unstable modes would be
shorter than the dynamical time of gravitational radiation) would give rise to a pinch-off similar to 
the one studied here. 

Finally, evolve unstable black strings (and other black objects) within 
asymptotically Anti de Sitter (AdS) spacetime, and explore the consequences within the
context of the AdS/Conformal Field Theory (CFT) dualities of string theory. This is partly related to 
the point mentioned above about further investigating the intriguing connections between gravity and
fluids, as certain states within a CFT will admit a hydrodynamic description. Though regardless,
understanding the CFT-duals to unstable black hole spacetimes is interesting in its own right,
and could have implications to the more recent applications of AdS/CFT to model certain condensed matter
and high energy particle physics systems.

\section{Acknowledgements}
We thank A. Buchel, V. Cardoso, M. Choptuik, R. Emparan, D. Garfinkle, K. Lake, S. Gubser, G.
Horowitz, D. Marolf, R. Myers, E. Poison, W. Unruh and R. Wald
for stimulating discussions. This work was supported by
NSERC (LL), CIFAR (LL), the Alfred P. Sloan Foundation (FP), and NSF grant PHY-0745779 (FP).
Research at Perimeter Institute is supported through Industry Canada and by the
Province of Ontario through the Ministry of Research \&
Innovation.

\bibliographystyle{cambridgeauthordate}
\bibliography{chapter3}

\end{document}